# The effects of gender, age and academic rank on research diversification[1]

Giovanni Abramo

*Laboratory for Studies in Research Evaluation*
*at the Institute for System Analysis and Computer Science (IASI-CNR)*
*National Research Council of Italy*
ADDRESS: Istituto di Analisi dei Sistemi e Informatica, Consiglio Nazionale delle Ricerche, Via dei Taurini 19, 00185 Roma - ITALY
giovanni.abramo@uniroma2.it

Ciriaco Andrea D'Angelo

*University of Rome "Tor Vergata" - Italy and*
*Laboratory for Studies in Research Evaluation (IASI-CNR)*
ADDRESS: Dipartimento di Ingegneria dell'Impresa, Università degli Studi di Roma "Tor Vergata", Via del Politecnico 1, 00133 Roma - ITALY
dangelo@dii.uniroma2.it

Flavia Di Costa

*Research Value s.r.l.*
ADDRESS: Research Value, Via Michelangelo Tilli 39, 00156 Roma - ITALY
flavia.dicosta@gmail.com

**Abstract**

In this work we analyze the combined effects of gender, age and academic rank on the propensity of individual scholars to diversify their scientific activity. The aspect of research diversification is measured along three main dimensions, namely its extent and intensity and the cognitive relatedness of the fields of diversification. We apply two regression models to the dataset of scientific output of all Italian professors in the sciences over the period 2004-2008. The aim is to understand how personal and organizational traits can influence individual behaviors in terms of research diversification. Among other considerations, our findings urge caution in identifying research diversification as a co-determinant of the gender productivity gap between males and females.

# 1. Introduction

We can expect researchers to engage differently in two behaviors or attitudes: leading either to strong concentration of research in one or few closely related fields, or to diversification of scientific activities among several fields. The challenges of complex societal problems requiring integration of differing competencies, as well as policy initiatives promoting interdisciplinary research, serve as incentives for scientists to apply their knowledge in fields other than their own.

Many prominent researchers seem to pursue a "scatter-gather" strategy, essentially remaining focused on one or two fields over the course of their career (Chakraborty, Tammana, Ganguly, & Mukherjee, 2015). Schuitema and Sintov (2017) consider that the researcher's own disciplinary competencies serve as their point of departure, with incursions into other fields taken as opportunities for the acquisition and offer of complementary competencies. Abramo, D'Angelo, and Di Costa (2017) studied diversification in research activities by Italian academics, along three dimensions: extent, intensity and cognitive relatedness of field diversification. They found that the extent of diversification (number of topics covered by the scientist's research portfolio, different than the dominant one) varies among fields within individual disciplines and among disciplines, and is highly correlated to the intensity of publication, being lowest in mathematics and highest in chemistry. Intensity of diversification (share of publications outside the academic's dominant topic of research) is lowest in earth sciences and highest in industrial and information engineering, but differences are generally less notable among disciplines. Degree of relatedness (share of publications falling in the same discipline) is lowest in earth sciences and highest in chemistry.

The same authors now intend to deepen their previous analysis, investigating the relation between extent, intensity and relatedness of field diversification, and the personal and organizational traits of the scientist's gender, age and academic rank.

There have been many studies investigating the issues of gender differences in research, however very few of these have examined the specific question of differences in the propensity to diversify. Leahey (2006) contends that in addition to institutional issues, the extent of research specialization can be a contributing factor to processes through which gender affects research productivity, and that the preference of women for less specialized research hinders their capacity to achieve publication and citation. In fact Rhoten and Pfirman (2007) find that female scientists are much more likely than males to: i) adapt tools, concepts, data, methods and results from different fields and/or disciplines; ii) collaborate in teams and networks that seek to exchange knowledge and tools across fields and/or disciplines; iii) undertake research in domains at intersections or on edges of multiple fields and/or disciplines; iv) engage in topics that not only draw on multiple fields and/or disciplines, but also serve multiple stakeholders and broader missions outside of academia. Abramo and D'Angelo (2017) demonstrated that gender impact on the extent of research diversification is meaningful both at overall level and at discipline level, with the sole exceptions of physics and earth sciences. Males tend to diversify research activity more than females, however this seems mainly due to their higher publication intensity. When controlled for individual output, gender differences appear significant in only two disciplines, mathematics and biology, and in the first one it is females that show higher extent of diversification, confirming the views of both Leahey (2006) and Rhoten and Pfirman (2007); conversely, in biology their views are contradicted. The greater propensity of females in mathematics to diversify is confirmed along another dimension, called "intensity of diversification", which also indicates higher intensity of diversification of women in the discipline of industrial and information engineering. Furthermore, males tend to diversify their research into more closely related fields, with this difference occurring both at the overall level and in the individual disciplines of physics and biology.

No studies have specifically investigated the relations of age and academic rank with research diversification, although there have been a few on age and rank versus collaboration and interdisciplinary research, from which we can get some hints. Bozeman & Corley (2004) note that over the course of their careers, tenured researchers develop a level of scientific and technical human capital, as well as social capital, which makes it possible for them to install ever more new collaborations. But the research achieved through interdisciplinary collaborations appears most attractive to those senior researchers who can assume the risks deriving from the highly unforeseeable results. In this regard, Rhoten (2004) asserts that while many young scientists might indeed be attracted by the intellectual rewards of interdisciplinary research, they can also be disincentivated by the professional risks posed to them as early tenure-track scientists. One of the perceived risks is that in moving away from their own disciplinary area, the researcher could prejudice their career progress, since recognition structures are typically discipline based. Abbot (2001) affirms that "the disciplinary system of departments is central to the careers and hiring of academics". In contrast, Millar (2013) argues that developing an interdisciplinary dissertation can increase the individual's possibility of obtaining an academic position, and that in the end interdisciplinary research does not have a dramatic effect on the types of positions individuals hold in academia.

Given the paucity of investigation and contrasting opinions, our intention with this work is to fill the void, by analyzing the combined effects of gender, age and academic rank on the propensity to diversify, along its three dimensions of extent, intensity, and field-relatedness. To this purpose, we follow the same approach as Abramo, D'Angelo, and Di Costa (2017), applied to the research production of all Italian professors in the sciences for the 2004-2008 period.

Applying an algorithm for disambiguation of author names developed by D'Angelo, Giuffrida and Abramo (2011), we are able to assign publications indexed in the Web of Science (WoS) to their Italian academic authors. Using the WoS classification scheme, each publication over the 2004-2008 period is then assigned to one or more subject categories (SCs), depending on the classification of the hosting journal. In this way we can assess the research diversification of each professor, and then, identifying their personal/organizational traits (gender, age, academic rank, field of research), we can pursue the research objective.

In the next section we illustrate the empirical methods of the study, in terms of indicators, dataset and statistical models. Section 3 presents the analytical results for each of the three dimensions of diversification. Finally, Section 4 concludes the work with the author's considerations.

## 2. Methodology and dataset

The following subsections provide a brief overview of the analytical methodology. The dataset and indicators are the same as those used in Abramo, D'Angelo and Di Costa (2017), so the reader will find a significant overlap with the like subsections of the preceding work.

### 2.1 Dataset

The dataset for the analysis is the 2004-2008 research production achieved by all Italian professors in the sciences. In the Italian academic system all professors are classified in one and only one field (named "scientific disciplinary sector", or SDS, of which 370 in all), grouped into disciplines (named "university disciplinary areas", UDAs, 14 in all).[2] In this study we focus on the sciences, for which the WoS coverage of publications by Italian universities is satisfactory. The sciences consist of 192 SDSs grouped into nine UDAs.[3]

Data on academics are extracted from a database maintained at the central level by the Ministry of Education, University and Research,[4] indexing the name, academic rank, affiliation, and the SDS of each professor. Publication data are drawn from the Italian Observatory of Public Research (ORP), a database developed and maintained by the authors and derived under license from the WoS. Beginning from the raw data of Italian publications indexed in the WoS, we apply a complex algorithm for disambiguation of the true identity of the authors and their institutional affiliations (for details see D'Angelo, Giuffrida, & Abramo, 2011). The final dataset is composed of 31,101 professors, specifically all and only those:
- operating in the 192 science SDSs;
- with at least one publication indexed in the WoS over the period 2004-2008;
- for which we can identify the individual's age as of 31/12/2008.[5]

Table 1 shows the division of professors by UDA, as well as the relative scientific production[6] for the observed five-year period.

### 2.2 Indicators

The disciplines in which research activity is classified often overlap and generally have quite weak boundaries – the fields within them even more so. The confines are in constant flux, as scientific progress continuously varies the scope of the fields and leads to the birth of new ones and disappearance of old ones. In spite of this, the analysis of research diversification requires some kind of classification. For our study we use the WoS system. We associate each publication in the database with only one topic. By "topic" we mean the WoS SC of the hosting journal in the case of a mono-category journal, or the combination of SCs when the publication is issued in a multi-category journal.

---

[2] For the complete list see http://attiministeriali.miur.it/UserFiles/115.htm, last accessed September 4, 2017.
[3] Mathematics and computer sciences, Physics, Chemistry, Earth sciences, Biology, Medicine, Agricultural and veterinary sciences, Civil engineering, Industrial and information engineering.
[4] See http://cercauniversita.cineca.it/php5/docenti/cerca.php, last accessed September 4, 2017.
[5] Ages were obtained by analysing an MIUR census from 2004, listing academics with voting rights for election of career-advancement selection committees. From this we identified the birth dates of 31,101 professors, authoring at least one WoS publication over the 2004-2008 period, representing 92.1% of the population in the 192 SDSs observed.
[6] Article, reviews, letters and conference proceedings.

The authors can thus be divided into two classes: i) those who diversify, meaning their publications fall in more than one topic; ii) those who do not diversify, meaning their publications fall in a single topic. We refer to these as "diversified" or "specialized" authors. Obviously the distribution of the scholars between the two classes depends on the breadth of the publication window, as well as the disciplinary classifications. The object of our study is the diversified authors. For each, we first identify the dominant topic in which the individual works, meaning the most recurrent SC or SC combination in their publication portfolio. We consider the case of a certain Mario Rossi (equivalent to a Jane or John Smith), professor in FIS/03 (Physics of matter), who in the period of observation produced eight articles published in four different journals (*Physical Review B*, *Physical Review E*, *Chemphyschem* and *Physical Review letters*). Given the classification of these journals under the WoS system, we have the distribution illustrated in Table 2. The eight articles fall in four different topics, of which the dominant one is subject category UK (Physics, condensed matter), given that half of Rossi's publications fall in this topic. Among other professors we observe cases of more than one dominant topic, which becomes particularly likely when the individual's number of publications is low.

We will investigate three dimensions of research diversification by individual professors. The first is "*extent of diversification (ED)*", measured by an indicator of the same name, given by the number of topics covered in the person's scientific portfolio. The second is intensity of diversification, meaning the share of the researcher's output falling outside their sector of specialization − measured by the indicator "*diversification ratio (DR)*", given by the ratio of the share of papers in topics other than the dominant one to total number of publications. The higher the value of ED and the closer DR is to one, the more the individual's research activity is diversified. The opposite situation denotes a highly specialized researcher. There can also be antithetical situations: i) a high ED value with low DR value would indicate that the subject is predominantly specialized but open to exploring new fields; ii) a low ED value with DR value tending to one is quite unlikely, unless the scientific production is very low. The last dimension investigated is the cognitive relatedness of the topics studied by the academic. Measurement of this requires definition of a threshold of proximity. For this purpose we associate the individual WoS topics to the disciplines,[7] meaning we can identify the topics as "related" if they fall within the same discipline. The indicator for this dimension is "*relatedness ratio (RR)*", equal to the ratio of number of papers in the dominant discipline to total number of papers. An RR of 1 indicates that the researcher, although diversifying, does not go beyond their own disciplinary area. It is likely that a statistician (whose sphere of research can range from statistics to economy, medicine,

---

[7] Each WoS subject category is associated with a single discipline, i.e., one of: Mathematics; Physics; Chemistry; Earth and space sciences; Biology; Biomedical research; Clinical medicine; Psychology; Engineering; Economics; Law, political and social sciences; Multidisciplinary sciences; Art and humanities.

agriculture, etc.) would have a much lower degree of relatedness than a surgeon.

According to the above taxonomy, for Mario Rossi we observe:
- an *ED* of 4;
- a *DR* of 0.5, since half of the total publications fall outside his dominant topic (Physics, condensed matter);
- an *RR* of 7/8, since 7 of the 8 publications are associated with the dominant discipline (Physics).

Note that given the different operating definition of the indicators, the three datasets relative to ED, DR, RR are of different sizes:

ED: 31,101 observations in all, i.e. all professors with at least one publication in the period examined;

DR: 26,942 observations in all, i.e. all professors with at least one publication falling in topics other than the dominant one;

RR: 23,007 observations in all, i.e. all professors with at least one publication falling in disciplines other than the dominant one.

*Table 1: Dataset of the analysis*

| UDA | SDSs | Professors | Publications |
|---|---|---|---|
| 1 - Mathematics and computer science | 10 | 2,685 | 14,741 |
| 2 - Physics | 8 | 2,416 | 24,004 |
| 3 - Chemistry | 12 | 3,101 | 24,407 |
| 4 - Earth sciences | 12 | 1,004 | 4,701 |
| 5 - Biology | 19 | 4,729 | 28,097 |
| 6 - Medicine | 50 | 9,163 | 56,533 |
| 7 - Agricultural and veterinary sciences | 30 | 2,504 | 10,516 |
| 8 - Civil engineering | 9 | 1,104 | 4,435 |
| 9 - Industrial and information engineering | 42 | 4,395 | 33,816 |
| Total | 192 | 31,101 | 179,506* |

*\* The total is less than the sum of column data due to multiple counting of publications co-authored by professors of more than one UDA.*

*Table 2: Publication portfolio of a professor in the dataset*

| Topic | Discipline | No. of publications | WoS_ID |
|---|---|---|---|
| UK (Physics, condensed matter) | Physics | 4 | 243195800122; 245330200070; 260574500061;251986500011 |
| UF+UR (Physics, fluids & plasmas; Physics, mathematical) | Physics | 2 | 228818200106; 242408800041 |
| EI+UH (Chemistry, physical; Physics, atomic, molecular & chemical) | Chemistry; Physics | 1 | 231971100043 |
| UI (Physics, multidisciplinary) | Physics | 1 | 229700800052 |

## 2.3 Statistical analysis

We apply two regression models to study the influence of the scientists' personal/organizational traits on each of our three indicators of research diversification. In the first (Model 1) we assess the relations between the value of each indicator and the professors' traits via a direct OLS regression. For extent of diversification ED, the model is formulated:

$$\text{ED} = a_0 + \sum_{n=1}^{4}(a_n \cdot \text{x}_n)$$

[1]

in which: $x_1$ = age; $x_2$ = gender (1=male, 0=female); $x_3$ = academic rank_1 (1 if full professor, 0 otherwise); $x_4$ = academic rank_2 (1 if associate professor, 0 otherwise).[8]

In the second model (Model 2) the dependence of the extent (as well as of intensity and relatedness) of field diversification on personal traits of the scientist is controlled for the number of publications authored by the professor. In fact, the literature clearly indicates that publication intensity varies with personal/organizational traits, i.e.:

- with gender (see Abramo, D'Angelo, & Caprasecca, 2009; Mauleón & Bordons, 2006; Xie & Shauman, 1998; Long, 1992; Cole & Zuckerman, 1984; Fox, 1983),
- with age (see Abramo, D'Angelo, & Murgia, 2016; Lissoni, Mairesse, Montobbio, & Pezzoni, 2011; Costas, van Leeuwen, & Bordons, 2010; Shin & Cummings, 2010; Gonzalez-Brambila & Veloso, 2007; Levin & Stephan, 1989; Over, 1982),
- with academic rank (see Abramo, D'Angelo, & Di Costa, 2011; Turner & Mairesse, 2005; Blackburn, Behymer, & Hall, 1978).

So, with respect to equation [1], we add an independent variable ($x_5$), obtaining:

$$ED = a_0 + \sum_{n=1}^{5}(a_n \cdot x_n)$$

[2]

with $x_1, \ldots x_4$ = same as above; $x_5$ = total number of publications authored by the professor.

To account for the differing intensity of publication across SDSs, both the values of the dependent variable and of $x_5$ are normalized with respect to the average of the distribution for the given SDS.

## 3. Analysis and results

### 3.1 Extent of diversification

The first analysis concerns extent of diversification (ED), i.e. the number of topics covered in each professor's scientific portfolio. Table 3 shows the descriptive statistics for this indicator (dependent variable) and the personal traits of all professors in the dataset (independent variables), by UDA.

Of the 31,101 professors in this particular dataset, women represent 29% of total. Their incidence approaches that of men only in Biology (48.2%); in Chemistry the share drops to 38.4%, and furthest of all, to below 20%, in Engineering and in Physics. The average age overall is 51.1, with a minimum of 47.2 in Industrial and information engineering and a maximum of 53.5 in Physics. The minimum age of professors per UDA fluctuates between 27 and 29, and the maximum between 76 and 79. The standard deviations are practically identical, demonstrating almost equal age distributions among the UDAs. Concerning academic rank, the population is divided almost equally between the three possible ranks, which a slight prevalence of assistant professors (37.7%) over associate (30.5%) and full professors (31.8%). These shares remain similar among the different UDAs.

---

[8] The academic ranks of professors are recorded as of 31/12/2008. The three-category classification (full, associate, assistant professor) is represented by two dummies. Assistant professors are coded 0 for both regressors, serving as a baseline category to which the other two ranks are compared.

*Table 3: Descriptive statistics of extent of diversification (ED) and personal traits of professors, by UDA (2004-2008 data)*

| UDA* | Obs | Age | | | Gender | Academic rank | | ED | | | |
| --- | --- | --- | --- | --- | --- | --- | --- | --- | --- | --- | --- |
| | | Mean | Std Dev. | Min-Max | Females (%) | Full prof. (%) | Associate prof. (%) | % Professors diversifying | Mean | Std Dev. | Min-Max |
| 1 | 2,685 | 48.1 | 10.3 | 28-76 | 31.1% | 35.5% | 30.5% | 79.8% | 3.185 | 3.574 | 0-38 |
| 2 | 2,416 | 53.5 | 10.8 | 27-78 | 17.3% | 33.6% | 35.0% | 92.3% | 5.766 | 4.907 | 0-58 |
| 3 | 3,101 | 51.1 | 11.3 | 28-77 | 38.4% | 30.0% | 33.5% | 94.0% | 5.801 | 4.857 | 0-46 |
| 4 | 1,004 | 52.2 | 10.1 | 29-78 | 25.3% | 33.8% | 32.4% | 82.3% | 2.673 | 2.421 | 0-18 |
| 5 | 4,729 | 51.9 | 9.9 | 29-78 | 48.2% | 31.5% | 28.5% | 90.8% | 5.063 | 4.637 | 0-42 |
| 6 | 9,163 | 53.4 | 8.9 | 28-79 | 26.8% | 27.2% | 30.4% | 84.5% | 4.748 | 4.871 | 0-53 |
| 7 | 2,504 | 49.5 | 9.6 | 28-77 | 32.9% | 33.8% | 28.6% | 81.2% | 3.193 | 3.071 | 0-23 |
| 8 | 1,104 | 48.5 | 10.2 | 29-77 | 15.9% | 37.6% | 29.9% | 76.7% | 2.779 | 3.148 | 0-35 |
| 9 | 4,395 | 47.2 | 10.4 | 28-79 | 13.4% | 36.6% | 29.2% | 89.0% | 5.457 | 5.051 | 0-61 |
| Overall | 31,101 | 51.1 | 10.2 | 27-79 | 29.0% | 31.8% | 30.5% | 86.6% | 4.683 | 4.630 | 0-61 |

\* 1, Mathematics and computer science; 2, Physics; 3, Chemistry; 4, Earth sciences; 5, Biology; 6, Medicine; 7, Agricultural and veterinary sciences; 8, Civil engineering; 9, Industrial and information engineering

As for the dependent variable ED (last four columns of Table 3), we note that overall 86.6% of professors diversify their scientific activity in the five-year period under observation. At the disciplinary level, Chemistry is the UDA with the maximum share of such professors (94.0%), followed by Physics (92.3%) and Biology (90.8%). On the other hand are Civil Engineering and Mathematics and computer science, where over 20% of professors do not diversify. Professors in Industrial and information engineering show the highest extent of diversification: on average, professors in this UDA publish in 5.5 different topics, almost twice as many as in Earth sciences (2.7).

Table 4 presents the results from the OLS regression [1] for ED versus personal traits. We immediately see that age impacts significantly on extent of diversification, both overall and in each UDA. The negative coefficient for the regressor demonstrates that young academics tend to diversify their research to a greater extent.

*Table 4: OLS regression of ED vs personal traits, by UDA*

| UDA | Obs | Age | Gender | Rank_Full | Rank_Assoc | $R^2$ |
| --- | --- | --- | --- | --- | --- | --- |
| 1 | 2,685 | -0.015*** | 0.086*** | 0.512*** | 0.232*** | 0.048 |
| 2 | 2,416 | -0.024*** | 0.016 | 0.648*** | 0.292*** | 0.079 |
| 3 | 3,101 | -0.023*** | 0.083*** | 0.940*** | 0.428*** | 0.132 |
| 4 | 1,004 | -0.023*** | 0.013 | 0.648*** | 0.390*** | 0.085 |
| 5 | 4,729 | -0.019*** | 0.140*** | 0.820*** | 0.373*** | 0.133 |
| 6 | 9,163 | -0.020*** | 0.097*** | 0.868*** | 0.366*** | 0.136 |
| 7 | 2,504 | -0.015*** | 0.045 | 0.466*** | 0.280*** | 0.043 |
| 8 | 1,104 | -0.024*** | 0.105* | 0.697*** | 0.317*** | 0.076 |
| 9 | 4,395 | -0.024*** | 0.040 | 0.716*** | 0.456*** | 0.084 |
| Overall | 31,101 | -0.018*** | 0.067*** | 0.693*** | 0.339*** | 0.089 |

*OLS estimation method. Dependent variable: Extent of diversification (ED). Independent variables: personal traits (for gender, F=0, M=1). Statistical significance: \*p-value <0.10, \*\*p-value <0.05, \*\*\*p-value <0.01. "Obs" is the number of professors with at least one publication in the period examined.*

Concerning the impact of gender, all coefficients are positive, but we see significance in only five out of nine UDAs (as well as overall). In Mathematics and computer science, Chemistry, Biology, Medicine, Civil Engineering, men show higher values of ED than women. The greatest marginal effect is in Biology, where the average gender difference for ED is 0.14.

Concerning impact of academic rank, comparing with the reference category of assistant professors, both full and associate professors show positive and statistically significant coefficients for difference in ED, in all UDAs and overall. Furthermore, coefficients for full professors are consistently higher than those for associates, leading to the conclusion that higher academic rank corresponds to greater extent of diversification in research activity.

The values of $R^2$ (proportion of variation in the dependent variable explained by the explanatory variables) are very low, with a maximum of only 0.136 in Medicine. However these values increase notably when we move to Model 2, which introduces the additional explanatory variable of the number of publications authored by each professor (i.e. total output). Table 5 shows the results of OLS regressions carried out applying ED equation [2]. In this case, all values of $R^2$ are greater than 0.6, with the sole exception of UDA 2 (Physics).

*Table 5: OLS regression of ED vs personal traits and total output, by UDA*

| UDA | Obs | Age | Gender | Rank_Full | Rank_Assoc | Total output | $R^2$ |
|---|---|---|---|---|---|---|---|
| 1 | 2,685 | 0.002 | -0.027 | -0.091** | -0.069** | 0.696*** | 0.626 |
| 2 | 2,416 | -0.019*** | -0.001 | 0.498*** | 0.231*** | 0.189*** | 0.185 |
| 3 | 3,101 | -0.007*** | 0.003 | 0.188*** | 0.106*** | 0.651*** | 0.647 |
| 4 | 1,004 | -0.002 | -0.032 | 0.045 | 0.010 | 0.668*** | 0.626 |
| 5 | 4,729 | -0.004*** | 0.007 | 0.155*** | 0.086*** | 0.669*** | 0.756 |
| 6 | 9,163 | -0.005*** | -0.011 | 0.178*** | 0.102*** | 0.567*** | 0.680 |
| 7 | 2,504 | 0.000 | -0.026 | 0.019 | -0.011 | 0.698*** | 0.670 |
| 8 | 1,104 | -0.005*** | -0.043 | 0.116*** | 0.048 | 0.671*** | 0.784 |
| 9 | 4,395 | -0.001 | -0.024 | 0.025 | 0.032* | 0.705*** | 0.735 |
| Overall | 31,101 | -0.004*** | -0.011** | 0.153*** | 0.086*** | 0.576*** | 0.612 |

*OLS estimation method. Dependent variable: Extent of diversification (ED). Independent variables: personal traits (for gender, F=0, M=1) and total output. Statistical significance: \*p-value <0.10, \*\*p-value <0.05, \*\*\*p-value <0.01. "Obs" is the number of professors with at least one publication in the period examined.*

Under Model 2, the coefficients for age are statistically significant and negative at the overall level and in five of nine UDAs (Physics, Chemistry, Biology, Medicine; Civil engineering). For these disciplines, including when we control for publication intensity, we therefore confirm that young academics tend to diversify their research activity to greater extent, although with marginal effects on the dependent variable (limited to the third decimal place, second decimal place in Physics). Concerning the influence of gender, the coefficient at overall level is significant but negative, seeming to overturn the results from Model 1: controlling for publication intensity, it is the women that, other traits equal, now show greater extent of diversification. However none of the coefficients at UDA level result as significant, meaning that we are left with no firm conclusions concerning the effect of gender on extent of diversification.

Finally, concerning effects from academic rank, we observe that compared to the reference category of assistant professors, the coefficients for full professors are statistically significant both at the overall level and in six of nine UDAs (Mathematics and computer science, Physics, Chemistry, Biology, Medicine, Civil engineering). The coefficients are also consistently positive, with the sole exception of UDA 1 (Mathematics). The coefficients for associate professors confirm the significance of impact of academic rank on extent of research diversification in the six UDAs, and again confirm the unique position of Mathematics: in this UDA it is the assistant professors that show a marked tendency to diversification of research, when compared to their higher ranked colleagues.

## 3.2 Intensity of diversification

In this section we repeat the same analysis for the dependent variable DR, diversification ratio, i.e the share of papers falling in topics other than the dominant one out of total publications by the professor. Tables 6 and 7 show the results from the regression analyses under models 1 and 2.

*Table 6: OLS regression of DR vs personal traits, by UDA*

| UDA | Obs | Age | Gender | Rank_Full | Rank_Assoc | $R^2$ |
|---|---|---|---|---|---|---|
| 1 | 2,143 | 0.002** | -0.007 | -0.042** | -0.029 | 0.003 |
| 2 | 2,229 | -0.001 | 0.015 | 0.006 | 0.002 | 0.001 |
| 3 | 2,916 | -0.004*** | 0.016 | 0.120*** | 0.054*** | 0.016 |
| 4 | 826 | -0.002* | -0.011 | 0.051 | 0.025 | 0.004 |
| 5 | 4,294 | -0.003*** | 0.009 | 0.105*** | 0.063*** | 0.019 |
| 6 | 7,744 | -0.001*** | 0.008 | 0.039*** | 0.012 | 0.002 |
| 7 | 2,033 | -0.002** | 0.001 | 0.064*** | 0.019 | 0.005 |
| 8 | 847 | -0.003*** | -0.001 | 0.105*** | 0.055** | 0.016 |
| 9 | 3,910 | -0.002*** | -0.014 | 0.064*** | 0.055*** | 0.009 |
| Overall | 26,942 | -0.002*** | 0.004 | 0.051*** | 0.026*** | 0.004 |

*Dependent variable: Diversification Ratio (DR). Independent variables: personal traits (for gender, F=0, M=1). OLS estimation method. Statistical significance: \*p-value <0.10, \*\*p-value <0.05, \*\*\*p-value <0.01. "Obs" equals the number of professors with at least one publication falling in topics other than the dominant one.*

Under both models we observe significant overlap with the results from the ED analysis, which we would certainly expect given the correlation between the two indicators.

Under Model 1 (Table 6), age impacts DR negatively, both overall and at UDA level, with the sole exception of Mathematics. On the contrary, the impact of academic rank is positive: professors of higher rank show significantly greater intensity of research diversification compared to their colleagues, both at overall level and in the large part of individual UDAs. Mathematics is again an anomaly: in comparison with colleagues of lower rank, the full professors show significantly lower DR.

However, under Model 1, the values of $R^2$ are remarkably low, indicating the scarce capacity of personal traits alone to explain the observed variance in intensity of research diversification for our population. In addition, the variance explained by the regressions increases very little under Model 2, meaning with the added covariate of number of publications authored by the scientist (last column, Table 7). We see clearly that number of publications is correlated with diversification intensity (coefficients are all positive and significant, except for Physics), but unlike the analysis for extent of diversification, the variable does not have much bearing on the relation between the professor's personal traits and their diversification behavior measured through DR.

*Table 7: OLS regression of DR vs personal traits and total output, by UDA*

| UDA | Obs | Age | Gender | Rank_Full | Rank_Assoc | Total output | $R^2$ |
|---|---|---|---|---|---|---|---|
| 1 | 2,143 | 0.003*** | -0.018 | -0.090*** | -0.053*** | 0.070*** | 0.035 |
| 2 | 2,229 | -0.001 | 0.017 | 0.017 | 0.007 | -0.016*** | 0.006 |
| 3 | 2,916 | -0.003*** | 0.007 | 0.046** | 0.022 | 0.072*** | 0.054 |
| 4 | 826 | -0.001 | -0.014 | 0.011 | -0.001 | 0.059*** | 0.024 |
| 5 | 4,294 | -0.002*** | -0.005 | 0.037*** | 0.034*** | 0.076*** | 0.089 |
| 6 | 7,744 | -0.001 | 0.004 | 0.013 | 0.002 | 0.025*** | 0.009 |
| 7 | 2,033 | 0.000 | -0.007 | 0.014 | -0.011 | 0.091*** | 0.062 |
| 8 | 847 | -0.002 | -0.016 | 0.052* | 0.033 | 0.083*** | 0.103 |
| 9 | 3,910 | 0.000 | -0.021** | -0.006 | 0.014 | 0.080*** | 0.080 |
| Overall | 26,942 | -0.001*** | -0.001 | 0.014** | 0.010** | 0.045*** | 0.025 |

*Dependent variable: Diversification Ratio (DR). Independent variables: personal traits (for gender, F=0, M=1) and total output. OLS estimation method. Statistical significance: \*p-value <0.10, \*\*p-value <0.05, \*\*\*p-value <0.01. "Obs" equals the number of professors with at least one publication falling in topics other than the dominant one.*

At overall level, the Model 2 regression shows significant coefficients for all traits except gender. In the analysis at UDA level, the coefficients for age are significant only in Mathematics and computer science, Chemistry and Biology. Chemistry and Biology also share the same sign of coefficient: in these disciplines the intensity of diversification seems higher for young scientists, and also for full professors compared to the lesser ranks. Mathematics again presents the singular and significant situation of the exact opposite case: young and assistant professors seem more focused on their core topics compared to older and full professors.

### 3.3 Relatedness of field diversification

Our final analysis concerns the "degree of relatedness" of field diversification, measured by the RR indicator, i.e. ratio of number of papers in the dominant discipline to total number of papers authored by the professor. Independently of extent and intensity of research diversification, the closer RR is to 1 the greater will be the cognitive proximity of the scholar's research results.

Table 9 shows the results from OLS regression under Model 1 for RR. At the overall level (last line of the table), the professors' personal traits seem to impact significantly on the RR indicator: relatedness of research diversification decreases with both increasing age and academic rank[9] and is also higher for the male professors of the observed population.

However, at the UDA level, the significance of the OLS coefficients is remarkably inconsistent. The results for Mathematics are the same as the overall case, with the exception of no significance for the gender regressor. In fact, concerning gender, only Physics and Biology show significant and positive coefficients. Apart from

---

[9] Note that since this is a cross-section analysis, we cannot identify whether the dependent variable is increasing (or decreasing) in relation to the individual's age or to their career progression.

Mathematics, the effect of age on degree of relatedness is also significant and negative in Medicine and in Industrial and information engineering. These two UDAs are further alike in showing positive and significant impact from academic rank on degree of research relatedness.

*Table 8: OLS regression of RR vs personal traits, by UDA*

| UDA | Obs | Age | Gender | Rank_Full | Rank_Assoc | $R^2$ |
|---|---|---|---|---|---|---|
| 1 | 1,600 | -0.006*** | 0.004 | 0.141*** | 0.074*** | 0.035 |
| 2 | 1,911 | -0.001 | 0.043** | 0.012 | 0.016 | 0.003 |
| 3 | 2,713 | 0.000 | 0.005 | -0.037 | -0.007 | 0.001 |
| 4 | 460 | -0.001 | -0.013 | 0.082* | 0.049 | 0.012 |
| 5 | 3,919 | 0.000 | 0.022** | -0.003 | 0.006 | 0.001 |
| 6 | 6,784 | -0.001** | -0.004 | 0.046*** | 0.034*** | 0.004 |
| 7 | 1,509 | -0.002* | 0.025* | 0.025 | 0.024 | 0.004 |
| 8 | 764 | -0.001 | 0.042 | -0.044 | -0.070** | 0.014 |
| 9 | 3,347 | -0.004*** | 0.006 | 0.071*** | 0.023* | 0.012 |
| Overall | 23,007 | -0.001*** | 0.009** | 0.029*** | 0.020*** | 0.002 |

*Dependent variable: Degree of relatedness (RR). Independent variables: personal traits (for gender, F=0, M=1). OLS estimation method. Statistical significance: *p-value <0.10, **p-value <0.05, ***p-value <0.01. "Obs" equals the number of professors with at least one publication falling in disciplines other than the dominant one.*

Introducing number of publications as a covariate (Model 2), we obtain the OLS estimations shown in Table 10. There are no changes at overall level, meaning that all personal traits maintain significant impact on the dependent variable RR, with the same sign of coefficient as in Model 1, although at slightly lower values. At the UDA level the picture again remains similar to Model 1: Mathematics, Medicine and Industrial and information engineering show significant effects from age and academic rank, while in Physics and Biology we see significant effects from gender. Finally, values of $R^2$ are consistently very low, confirming that for this indicator, personal traits explain only a marginal part of the overall variance.

*Table 9: OLS regression of RR vs. personal traits and total output, by UDA*

| UDA | Obs | Age | Gender | Rank_Full | Rank_Assoc | Total output | $R^2$ |
|---|---|---|---|---|---|---|---|
| 1 | 1,600 | -0.006*** | -0.003 | 0.115*** | 0.060*** | 0.038*** | 0.046 |
| 2 | 1,911 | 0.001 | 0.036* | -0.036 | -0.002 | 0.074*** | 0.099 |
| 3 | 2,713 | 0.000 | 0.010 | 0.006 | 0.012 | -0.043*** | 0.013 |
| 4 | 460 | 0.001 | -0.015 | 0.039 | 0.021 | 0.067*** | 0.039 |
| 5 | 3,919 | 0.000 | 0.028*** | 0.022 | 0.017 | -0.030*** | 0.008 |
| 6 | 6,784 | -0.001* | -0.006 | 0.030*** | 0.028*** | 0.016*** | 0.007 |
| 7 | 1,509 | -0.001 | 0.023* | 0.016 | 0.018 | 0.016* | 0.007 |
| 8 | 764 | -0.001 | 0.044 | -0.038 | -0.068** | -0.009 | 0.015 |
| 9 | 3,347 | -0.004*** | 0.004 | 0.063*** | 0.018 | 0.010 | 0.013 |
| Overall | 23,007 | -0.001*** | 0.007* | 0.019*** | 0.015*** | 0.013*** | 0.004 |

*Dependent variable: Degree of relatedness (RR). Independent variables: personal traits (for gender, F=0, M=1) and total output. OLS estimation method. Statistical significance: *p-value <0.10, **p-value <0.05, ***p-value <0.01. "Obs" equals the number of professors with at least one publication falling in disciplines other than the dominant one.*

## 4. Conclusions

Like all explorers, researchers are subject to a natural sense of curiosity, stimulating them to enter little-investigated fields. There are also attractive opportunities for individual scholar to apply their competencies in areas different from their specialization, thus bringing about diversification in their research activities. The analysis of level of specialization, or conversely of diversification of research activities, has only recently been addressed in the literature, with the introduction of indicators based on disciplinary classification of the scientific output authored by individual scientists. In a preceding work, Abramo, D'Angelo and Di Costa (2017) carried out an analysis of the distribution of such indicators within and between scientific domains. In this paper, the same authors have investigated the influence of the researchers' personal attributes of age, gender and academic rank on their own research diversification.

The results show that age impacts significantly on extent of diversification (ED), with coefficients almost always negative, indicating that compared to older scientists, younger ones have a greater propensity to diversify their research activity. This result is confirmed by analysis of the second indicator, DR, or incidence of diversified works: the share of publications falling in topics other than the dominant one. Again, this indicator is greater for young scientists than for older ones.

Analyzing the effect of academic rank, it emerges that compared to colleagues of lower rank, full professors show a remarkable tendency to increase the number of their research topics. On controlling for scholars' total output this relation seems diminished, but it remains clear that those in earlier career stages tend to specialize their research more than do full professors. In effect, given their accumulated experience and reputation, higher-ranking professors tend to be more engaged in managing research groups, raising funds, and in general in strategic external relations. This could well determine a lesser focus on their core research activities compared to the situation for their colleagues of lower rank. This part of our findings seem to confirm the claims of some scholars that in their early career stages, researchers tend to remain within their own disciplinary areas, due to the negative potential effects of diversification (Bozeman & Corley, 2004; Rhoten, 2004; Abbott, 2001). But what emerges is then an apparent contradiction between the results for the age and academic rank regressors. In fact career advancement is directly correlated with age (Over, 1982; Bayer & Dutton, 1977); yet in our analysis, while specialist focus increases with the age regressor, on the other hand we also find that career advancement corresponds with greater breadth in the individual's portfolio of scientific topics. However, it should be clear that our analysis is not longitudinal: the cross-section dataset includes both young professors and older assistant professors. Therefore, the contradictory results concerning the age/career-stage variables could be explained in the light of a "cohort" effect, which would determine incentives (or disincentives) to research diversification depending on the combination of these variables, independently of their single correlations.

Lastly, we come to the effect of the gender regressor on the three indicators of extent, intensity and relatedness of diversification, where we see that the impact of this factor seems significant only for extent of diversification, and only if not controlled for publication intensity. We could therefore conclude that there are no gender differences concerning the propensity to focus or diversify individual research. In fact, the different research diversification behavior of females with respect to males seems to vary based on the dimension of diversification analyzed, and on the specificity of each discipline.

This induces us to recommend caution in identifying research diversification as a co-determinant of the gender productivity gap between males and females.

The results of all our analyses depend to some extent on both the publication window selected (for which it would be interesting to conduct a sensitivity analysis) and the field classification schemes of scientific output and of the professors themselves. In particular, the current paper deals with the Italian academic system, and the replicability of the analyses and validation of results would require development of similar classification schemes for other large-scale scientific communities.